\documentstyle[12pt,A4,epsfig,axodraw]{article}
\input epsfig.sty

\begin{document}

\begin{center}
{\LARGE\bf Heavy flavour contributions to the spin structure function $g_1(x,Q^2)$ and the Bj{\o}rken sum rule}

\vspace{1cm}
{Author W.L. van Neerven \footnote{Talk presented at the Workshop on "Deep
Inelastic Scattering off Polarized Targets: Theory meets Experiment",DESY
Zeuthen, Germany, September 1-5, 1997}}

\vspace*{1cm}
{\it Instituut-Lorentz, University of Leiden, P.O. Box 9506, 2300 RA
Leiden, The Netherlands}\\

\vspace*{2cm}

\end{center}

\begin{abstract}
We discuss the order $\alpha_s^2$ corrections to the longitudinal 
spin structure function
$g_1(x,Q^2,m^2)$ which are due to heavy flavour contributions.
Here $Q$ denotes the virtuality of the photon and $m$ stands for
the heavy flavour mass. Since the exact heavy quark coefficient functions are 
not known yet we have used the asymptotic forms which are strictly speaking  
only valid in the region $Q^2 \gg m^2$. However an analysis of the exact
and asymptotic expressions for $F_2^{\rm NLO}(x,Q^2,m^2)$ and 
$g_1^{\rm LO}(x,Q^2,m^2)$ 
reveals that the asymptotic forms can be also used at smaller $Q^2$. It
appears that for the region $0.01 < x \leq 0.1$ the NLO charm quark component
can become twice as large as in LO. However it is still much smaller
than $g_1(x,Q^2)$ due to light parton contributions. This is in contrast to
the observations made for $F_2(x,Q^2)$ at small $x$.  Also the charm quark 
contribution to the Bj{\o}rken sum rule turns out to be very small.

\end{abstract}

\section{Introduction}
Heavy quark production in deep inelastic electron-proton scattering proceeds
via the following reaction (see Fig. \ref{fig:1})
\begin{eqnarray}
\label{eqn:1}
e^-(l_1) + P(p) \rightarrow e^-(l_2) + Q (p_1)\,( \bar Q(p_1) ) +'X'\,.
\end{eqnarray}
Where $'X'$ denotes any inclusive hadronic final state and $V$ stands for the
neutral intermediate vector bosons given by $\gamma,Z$. In this paper we 
treat the case that $Q^2 \ll M_Z^2$ so that the process in Fig. \ref{fig:1}
is dominated by the one photon exchange mechanism only.
If we also integrate
over the momentum $p_1$ of the heavy (anti) quark $Q$ the unpolarised cross
section is given by
\begin{eqnarray}
\label{eqn:2}
\frac {d^2\sigma}{dx\,dy}=\frac{2 \pi \alpha^2}{(Q^2)^2} S_{eP}
 \Big[ \{ 1 + (1-y)^2\}
F_2(x,Q^2,m^2)- y^2 F_L(x,Q^2,m^2) \Big]\,.
\end{eqnarray}
Here $S_{eP}$ denotes the centre of mass energy of the electron-proton system
and the heavy quark component of the spin averaged structure functions are 
given by $F_i(x,Q^2,m^2)$ ($i=2,L$). In the case the proton is polarised
parallel ($\rightarrow$) or anti-parallel ($\leftarrow$) with respect to the 
spin of the incoming electron we obtain the cross section
\begin{eqnarray}
\label{eqn:3}
\frac {d^2\sigma(\rightarrow)}{dx\,dy}-\frac {d^2\sigma(\leftarrow)}{dx\,dy}
=\frac{8 \pi \alpha^2}{Q^2}\Big[ \{ 2 - y \} g_1(x,Q^2,m^2) \Big]\,, 
\end{eqnarray}
where $g_1(x,Q^2,m^2)$ denotes the heavy quark component of the longitudinal
spin structure function. Further we have defined the scaling variables
\begin{eqnarray}
\label{eqn:4}
x=\frac{Q^2}{2pq} \hspace*{2cm} y = \frac{pq}{pl_1} \hspace*{2cm} q^2=-Q^2 < 0
\,.
\end{eqnarray}
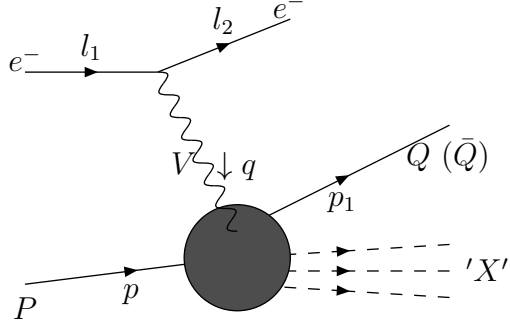
\begin{figure}
\begin{center}
  \begin{picture}(170,130)(0,0)
    \ArrowLine(0,20)(80,30)
    \ArrowLine(80,40)(160,80)
    \DashArrowLine(80,30)(160,35){5}
    \DashArrowLine(80,25)(160,25){5}
    \DashArrowLine(80,20)(160,15){5}
    \GCirc(80,30){20}{0.3}
    \Photon(50,100)(80,40){3}{7}
    \ArrowLine(0,100)(50,100)
    \ArrowLine(50,100)(100,120)
    \Text(0,110)[t]{$e^-$}
    \Text(100,130)[t]{$e^-$}
    \Text(0,15)[t]{$P$}
    \Text(160,75)[t]{$Q~(\bar Q)$}
    \Text(60,70)[t]{$V$}
    \Text(175,30)[t]{$'X'$}
    \Text(40,20)[t]{$p$}
    \Text(25,115)[t]{$l_1$}
    \Text(75,125)[t]{$l_2$}
    \Text(80,70)[t]{$\downarrow q$}
    \Text(120,55)[t]{$p_1$}
  \end{picture}
  \caption[]{Kinematics of charm production in deep inelastic
             electron-proton scattering}
  \label{fig:1}
\end{center}
\end{figure}
In QCD the twist two contributions to the charm component of the structure
functions defined above can be written as
\begin{eqnarray}
\label{eqn:5}
&& \frac{1}{n_f} \sum_{k=1}^{n_f} e_k^2 \int_x^{z_{max}} \frac{dz}{z}
\Big [ f_q^{\rm S}(\frac{x}{z},\mu^2) \otimes 
L_{i,q}^{\rm S}(z,\frac{Q^2}{m^2},\frac{m^2}{\mu^2})
\nonumber\\
&& + f_g^{\rm S}(\frac{x}{z},\mu^2) \otimes 
L_{i,g}^{\rm S}(z,\frac{Q^2}{m^2},\frac{m^2}{\mu^2})
 + n_f f_k^{\rm NS}(\frac{x}{z},\mu^2) \otimes L_{i,q}^{\rm NS}
(z,\frac{Q^2}{m^2},\frac{m^2}{\mu^2}) \Big ]
\nonumber\\
&& + e_Q^2 \Big [ f_q^{\rm S}(\frac{x}{z},\mu^2)
\otimes H_{i,q}^{\rm PS}(z,\frac{Q^2}{m^2},\frac{m^2}{\mu^2})
 + f_g^{\rm S}(\frac{x}{z},\mu^2) \otimes 
H_{i,g}^{\rm S}(z,\frac{Q^2}{m^2},\frac{m^2}{\mu^2})
\Big ] \,,
\end{eqnarray}
with
\begin{eqnarray}
\label{eqn:6}
z_{max}= \frac{Q^2}{Q^2+4m^2}\qquad \mbox{which follows from} \qquad
s=\frac{(1-z)}{z} Q^2 \geq 4m^2\,.
\end{eqnarray}
Further $n_f$ denotes the number of light flavours and $\mu$ stands for the
mass factorization scale which is put to be equal to the renormalization scale.
The charges of the light quark $k$ and the heavy quark $Q$ are given by $e_k$
and $e_Q$ respectively. 
Notice that $n_f$ is determined by all quarks which are lighter than
the heavy quark $Q$ and Eq. (\ref{eqn:5}) has to be understood in the fixed
flavour number scheme.
In Eq. (\ref{eqn:5}) the gluon density is denoted
by $f_g^{\rm S}(z,\mu^2)$ and the singlet (${\rm S}$) and non singlet 
(${\rm NS}$) combination of quark densities are defined by
\begin{eqnarray}
\label{eqn:7}
f_q^{\rm S}(z,\mu^2) & = & \sum_{k=1}^{n_f} [ f_k(z,\mu^2)+f_{\bar k}(z,\mu^2)]
\,,
\end{eqnarray}
\begin{eqnarray}
\label{eqn:8}
f_k^{\rm NS} (z,\mu^2) & = & f_k(z,\mu^2) + f_{\bar k}(z,\mu^2)
-\frac{1}{n_f}f_q^{\rm S}(z,\mu^2) \,.
\end{eqnarray}
The heavy quark coefficient functions $L_{i,k}$ and $H_{i,k}$ 
($i=1,2,L;~~k=q,g$)
are distinguished according to the
production mechanisms from which they originate. The quantity $L_{i,k}$ is 
given by the processes where the virtual photon couples to the light quark
whereas $H_{i,k}$ originates from the reactions where the virtual photon
is attached to the heavy quark. Hence these coefficient functions in 
Eq. (\ref{eqn:5}) are multiplied by $e_k^2$ and $e_Q^2$ respectively.

\begin{figure}
 \begin{center}
  {\unitlength1cm
  \epsfig{file=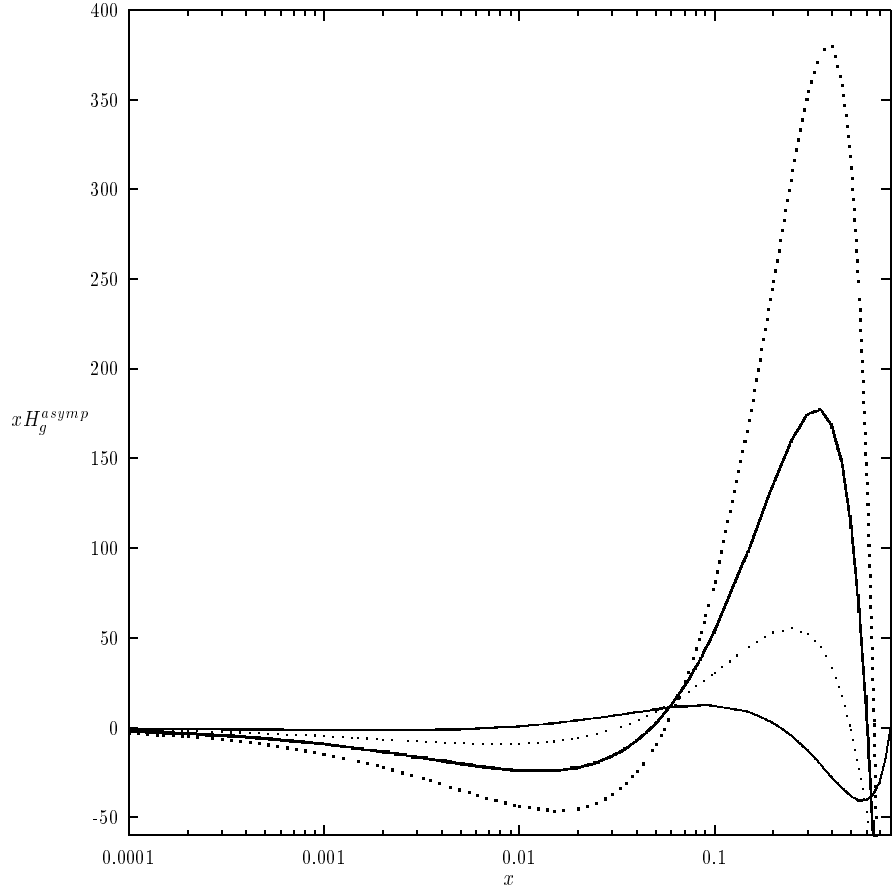,bbllx=52pt,bblly=272pt,bburx=536pt,bbury=710pt,%
    height=12cm,width=12cm,clip=,angle=0}
     }
\end{center}
\caption[]{The function $xH_{1,g}^{{\rm asymp},(2)}(x,\frac{Q^2}{m_c^2},1)$
($m_c=1.5~{\rm GeV/c}$) for $Q^2=10$
(lower solid line), $Q^2=100$ (lower dotted line),
$Q^2=10^3$ (upper solid line),
$Q^2=10^4$ (upper dotted line). All units are in $({\rm GeV/c})^2$}.
  \label{fig:2}
\end{figure}

Up to order $\alpha_s^2$ the heavy quark coefficient functions are given by 
the following parton subprocesses.
In order $\alpha_s$ we have the virtual photon gluon fusion mechanism given by
\begin{eqnarray}
\label{eqn:9}
            \gamma^* + g \rightarrow Q + \bar Q\,,
\end{eqnarray}
{}from which one obtains the heavy quark coefficient function 
$H_{i,g}^{{\rm S},(1)}$.
Here the superscript denotes the order in the perturbation series. In next-to-
leading order (NLO) one has the following processes. First one has to include
the virtual gluon corrections to reaction (\ref{eqn:9}) which have to be
added to the gluon bremsstrahlung process
\begin{eqnarray}
\label{eqn:10}
\gamma^* + g \rightarrow Q + \bar Q + g\,.
\end{eqnarray}
This leads to the coefficient function $H_{i,g}^{{\rm S},(2)}$.
Finally we also have the subprocess
\begin{eqnarray}
\label{eqn:11}
\gamma^* + q(\bar q) \rightarrow Q + \bar Q + q(\bar q)\,.
\end{eqnarray}
The above reaction has two different production mechanisms. The first one is
the Bethe-Heitler process where the virtual photon is coupled to the heavy 
quark which leads to the coefficient function $H_{i,q}^{{\rm PS},(2)}$. 
Here PS means purely singlet since this function originates from the partonic
process where a gluon (flavour singlet) is exchanged in the $t$-channel between
the light quark $q$ and the heavy quark $Q$. The 
second one is the Compton process where the virtual photon is coupled to the
light quark from which one obtains $L_{i,q}^{{\rm NS},(2)}=
L_{i,q}^{{\rm S},(2)}$. The expressions for the order $\alpha_s$ contribution
$H_{i,g}^{{\rm S},(1)}$ can be found in \cite{Wit} ($i=L,2$; unpolarised)
and \cite{Wat} ($i=1$; polarised). The exact analytic expressions for 
$L_{i,q}^{{\rm NS},(2)}$ are calculated in \cite{Bmsmn} ($i=L,2$; unpolarised)
and \cite{Bmsn1} ($i=1$; polarised). The exact heavy quark coefficient 
functions $H_{i,q}^{{\rm PS},(2)}$ and $H_{i,g}^{{\rm S},(2)}$ have been only 
calculated in the unpolarised case ($i=L,2$) (see \cite{Lrsn}) but the exact
expressions in polarised scattering ($i=1$) are not known yet. However in
\cite{Bmsmn} and \cite{Bmsn1} we could derive analytic formulae for the
asymptotic heavy quark coefficient functions up to order $\alpha_s^2$
for unpolarised and polarised electroproduction respectively.
The asymptotic heavy quark coefficient functions are defined by
\begin{eqnarray}
\label{eqn:12}
H_{i,k}^{\rm asymp}(x,\frac{Q^2}{m^2},\frac{m^2}{\mu^2})= \lim_{Q^2 \gg m^2}
\Big[H_{i,k}^{\rm exact}(x,\frac{Q^2}{m^2},\frac{m^2}{\mu^2})\Big] \,.
\end{eqnarray}
These asymptotic expressions are very useful because
\begin{itemize}
\item[1] They provide us with a check on the exact calculation.
\item[2] The exact calculation of $H_{i,q}^{{\rm PS},(2)}$ and 
$H_{i,g}^{{\rm S},(2)}$ ($i=2,L$) is semi-analytic and they are only 
available in the
form of long computer programs (see \cite{Lrsn}). The latter show numerical
instabilties at $Q^2 \gg m^2$ which are remedied by using in this region
$H_{i,k}^{\rm asymp}$. 
\item[3] Further if these asymptotic heavy quark coefficient functions are
substituted in Eq. (\ref{eqn:5}) it turns out that 
(see \cite{Bmsn2}, \cite{Buza})
\begin{eqnarray}
\label{eqn:13}
 R_2(x,Q^2,m^2)=
\frac{F_{2,Q}^{\rm asymp}(x,Q^2,m^2)}{F_{2,Q}^{\rm exact}(x,Q^2,m^2)}
\,,
\end{eqnarray}
tends to unity as $Q^2 > 10~m^2$ and $x<0.01$. 
In the case of charm production ($Q=c$ and $m_c=1.5~{\rm GeV/c})$
this happens for $Q^2 > 20~({\rm GeV/c})^2$. This implies that above a rather
low value of $Q^2$ it does not make any difference whether we use the asymptotic
or the exact expressions for $H_{i,k}$.
\item[4] The asymptotic forms of the heavy quark coefficient functions also
play a major role in the derivation of the variable number flavour scheme
(VFNS) from the fixed flavour number sheme (FFNS) 
(see \cite{Bmsn2}, \cite{Acot})
\end{itemize}
The asymptotic heavy quark coefficient functions get the typical form
\begin{eqnarray}
\label{eqn:14}
H_{i,k}^{{\rm asymp},(l)}(x,\frac{Q^2}{m^2},\frac{m^2}{\mu^2}) \sim \alpha_s^l
\sum_{n+j\leq l} a_{nj}(x)
\ln^n\Big(\frac{\mu^2}{m^2}\Big)\ln^j\Big(\frac{Q^2}{m^2}\Big)\,,
\end{eqnarray}
with an analogous expression for $L_{i,k}^{\rm asymp}$.
There are two ways to determine Eq. (\ref{eqn:14}).
\begin{itemize}
\item[1] Evaluate the phase space integrals and the Feynman integrals in
the usual way for $Q^2 \gg m^2$.
\item[2] Use operator product expansion techniques and the theorem of mass
factorization.
\end{itemize}
The last method was applied in \cite{Bmsmn} and \cite{Bmsn1} to compute the
asymptotic forms of the heavy quark coefficient functions for unpolarised
and polarised scattering respectively.

\section{Order $\alpha_s^2$ contributions to $g_1(x,Q^2,m^2)$}
In this section we will analyse the charm component of the longitudinal 
spin structure function $g_1$.
To compute the latter in NLO the 
following coefficient functions are available: $H_{1,g}^{{\rm exact},(1)}$
(Born), $L_{1,q}^{{\rm exact},(2)}$ (Compton process), 
$H_{1,g}^{{\rm asymp},(2)}$ (gluon
bremsstrahlung), $H_{1,q}^{{\rm asymp},(2)}$ (Bethe-Heitler). Like in 
unpolarised scattering the gluonic coefficient functions indicated by
$H_{1,g}^{(l)}$ are the most important ones. Therefore deep inelastic 
electroproduction of charm quarks provide us with an excellent way to determine
the gluon density in polarised scattering (see Eq. (\ref{eqn:5})). However as we
will show below the charm component in the spin case is much smaller with
respect to the light parton contribution to the structure function as observed
for $F_2$ in unpolarised scattering. This is revealed by 
$H_{1,g}^{{\rm asymp},(2)}$ in Fig. \ref{fig:2} 
where we observe a strongly 
oscillatory behaviour which is in contrast to its unpolarised analogue 
shown in Fig. 8  of \cite{Neer}. Further it satisfies the sum rule
\begin{eqnarray}
\label{eqn:15}
\int_0^1 dx\,\, H_{1,g}^{(l)}(x,\frac{Q^2}{m^2},\frac{m^2}{\mu^2})=0\,,
\end{eqnarray}
which holds in any scheme. Notice that the Born contribution $H_{1,g}^{(1)}$
also satisfies the above relation and we may safely assume that Eq. 
(\ref{eqn:15})
holds in all orders for general $Q^2$ and $m^2$. Further it turns out that
the coefficient functions $H_{1,k}^{(l)}$ ($k=q,g$) are much smaller than
their unpolarised analogues. In particular the large logarithmic terms of the
type $\frac{ln^i x}{x}$ which are characteristic of the latter do not show up
in the spin case. The absence of these corrections and the oscillatory 
behaviour leads to much smaller charm quark contributions to $g_1$ than
observed for $F_2$ or $F_L$.
To give a better estimate of the order 
$\alpha_s^2$ corrections we have to improve the asymptotic form of the
coefficient functions $H_{1,g}^{{\rm asymp},(2)}$ and 
$H_{1,q}^{{\rm asymp},(2)}$ a little bit. In \cite{Bmsn1} we multiplied them
by $\sqrt(1-\frac{4m^2}{s})$ and added to $H_{1,g}^{{\rm asymp},(2)}$ the
soft gluon bremsstrahlung term which is universal (see Eqs. (5.1)-(5.3) in
\cite{Bmsn1}). These approximations were tested for $F_2^{\rm NLO}$ and 
$g_1^{\rm LO}$
and we found a fairly good agreement with the exact results as long as
$x<0.1$ and $Q^2> 10~({\rm GeV/c})^2$.
\newpage

\begin{figure}[htbp]
\hspace*{0.0cm}
\epsfig{file=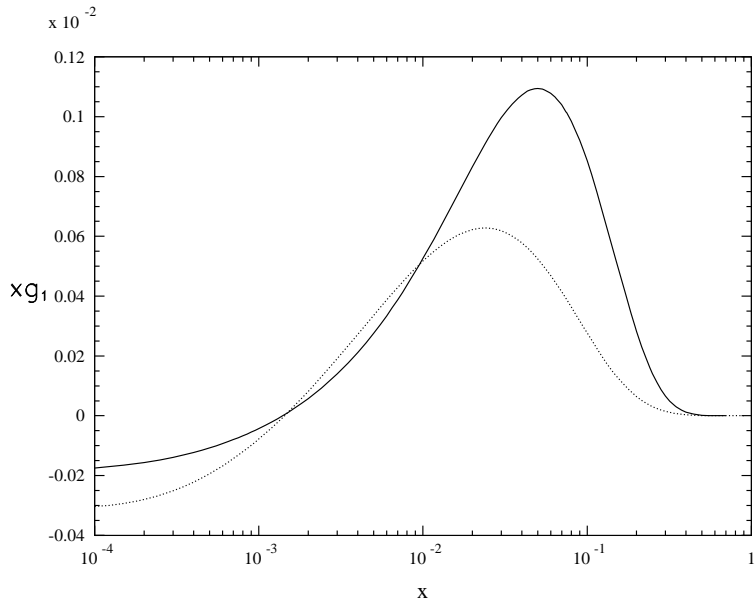,bbllx=-20pt,bblly=0pt,bburx=575pt,bbury=750pt,
        width=11.5cm,angle=-90}
\caption[]{The charm component of $xg_1(x,Q^2,m_c^2)$ at
$Q^2=50~({\rm GeV/c})^2$. Dotted line: $xg_1^{\rm exact}({\rm Born})$;
solid line: $xg_1^{\rm approx}({\rm NLO})$}.
  \label{fig:3}
\end{figure}

Choosing the parton density set in \cite{Grsv} (Standard Scenario) with $n_f=3$
and $\Lambda=200~{\rm MeV}$ we have plotted the charm component of 
$g_1^{exact}({\rm Born})$ and $g_1^{approx}({\rm NLO})$ at 
$Q^2=50~({\rm GeV/c})^2$ 
in Fig. 3. From this figure we infer that the charm
component becomes maximal in the region $0.01<x<0.1$. Further we observe that
in NLO the result becomes twice as large than in LO. We also made a comparison
in Fig. 4 with the quantity $g_1^{\rm light}({\rm NLO})$ which is due to light 
partons ($u,d,s$ and $g$) only. Here we see that the charm component amounts to
about $4 \%$ of $g_1^{\rm light}({\rm NLO})$. 
This is in contrast to the observation
made for the structure function $F_2$ in unpolarised scattering 
(see \cite{H1}, \cite{Zeus}). In the latter case the charm component becomes
large at very small $x$ (i.e. $x=10^{-4}$) and it amounts to about $25 \%$
with respect to the total structure function. In \cite{Bmsn1} we also studied
the spin structure functions plotted at $Q^2=10~({\rm GeV/c})^2$ and
$Q^2=100~({\rm GeV/c})^2$. These plots show the same features as the ones
made for $Q^2=50~({\rm GeV/c})^2$ so that our conclusions are unaltered.

\begin{figure}[htbp]
\vspace*{-1.2cm}
\epsfig{file=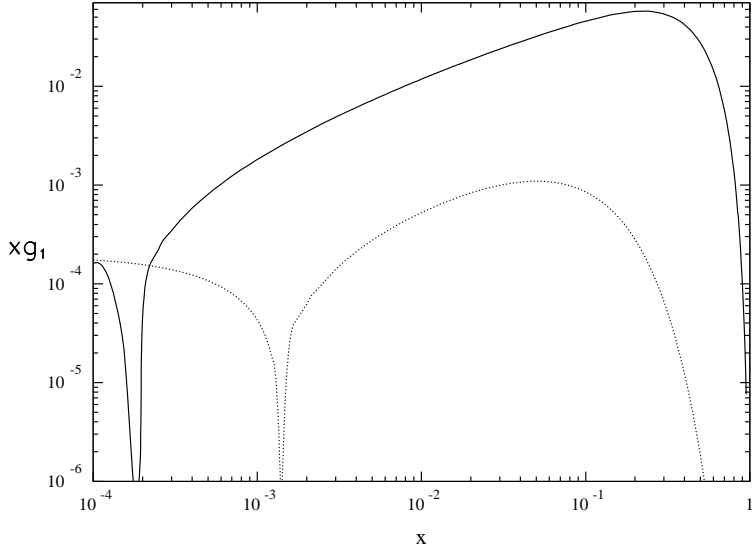,bbllx=-20pt,bblly=0pt,bburx=575pt,bbury=750pt,
        width=11.5cm,angle=-90}
\caption[]{The charm component and the light parton contribution to
$xg_1(x,Q^2,m_c^2)$ at $Q^2=50~({\rm GeV/c})^2$. Dotted line:
$xg_1^{\rm approx}({\rm charm,NLO})$; solid line: 
$xg_1^{\rm light}({\rm NLO})$}.
  \label{fig:4}
\end{figure}

\section{Order $\alpha_s^2$ contributions to the Bj{\o}rken sum rule}
\vspace{1mm}
\noindent
Since we know the exact result for $L_{1,q}^{{\rm NS},(2)}$ \cite{Bmsn1}
we can compute the order $\alpha_s^2$ correction to the Bj{\o}rken sum
rule due to heavy flavour contributions. The latter is given by
\begin{eqnarray}
\label{eqn:16}
\int_0^1 dx \Big [g_1^P(x,Q^2,m^2) - g_1^N(x,Q^2,m^2) \Big]= \frac{1}{6}
\mid \frac{G_A}{G_V} \mid \Big (L_{1,q}^{\rm NS} \Big )^{(1)}\,,
\end{eqnarray}
where the superscript $(1)$ denotes the first moment (Mellin transform).
In order $\alpha_s^2$ we have the following properties
\begin{eqnarray}
\label{eqn:17}
Q^2 \ll m^2 \hspace*{0.5cm} \rightarrow \hspace*{0.5cm}
\Big ( L_{1,q}^{{\rm NS},(2)}\Big )^{(1)} \sim \frac{Q^2}{m^2}\,,
\end{eqnarray}
which shows the decoupling of the heavy quark when $Q^2 \ll m^2$.
\begin{eqnarray}
\label{eqn:18}
Q^2 \gg m^2 \hspace*{0.5cm} \rightarrow \hspace*{0.5cm}
\Big ( L_{1,q}^{{\rm NS},(2)}\Big )^{(1)} = C_FT_f 
\Big (\frac{\alpha_s(n_f,\mu^2)}{4\pi}\Big )^2
\Big [ -4 \ln(\frac{Q^2}{m^2}) + 8 \Big]\,.
\end{eqnarray}
The light parton contribution to the Bj{\o}rken sum rule is given by
\begin{eqnarray}
\label{eqn:19}
\int_0^1 dx \Big [g_1^P(x,Q^2) - g_1^N(x,Q^2) \Big]= \frac{1}{6}
\mid \frac{G_A}{G_V} \mid \Big ({\cal C}_{1,q}^{\rm NS} \Big )^{(1)}\,,
\end{eqnarray}
where ${\cal C}_{1,q}^{\rm NS}$ denotes the massless quark coefficient 
function. The first moment of the latter quantity has been calculated up
to order $\alpha_s^3$ in \cite{Lave} and the order $\alpha_s^4$ term has been
estimated too (see e.g. \cite{Sam}. It turns out that the charm contribution
to the Bj{\o}rken sum rule is even smaller than the the estimated order
$\alpha_s^4$ correction (see \cite{Blum}). The contributions coming from the 
bottom and top quark are even smaller which can be attributed to the
decoupling of heavy quarks shown in Eq. \ref{eqn:17}. Finally we have the 
relation 
\begin{eqnarray}
\label{eqn:20}
{\cal C}_{1,q}^{\rm NS}(n_f,x,\frac{Q^2}{\mu^2})+
\lim_{Q^2 \gg m^2} L_{1,q}^{\rm NS}(x,\frac{Q^2}{m^2},\frac{m^2}{\mu^2})=
{\cal C}_{1,q}^{\rm NS}(n_f+1,x,\frac{Q^2}{\mu^2})\,,
\end{eqnarray}
so that for $Q^2 \gg m^2$ the large logarithm $\ln\frac{Q^2}{m^2}$ in the heavy 
quark coefficient function $L_{1,q}^{\rm NS}$ (Eq. (\ref{eqn:19}))
can be absorbed in the light
quark coefficient function ${\cal C}_{1,q}^{\rm NS}$. Hence the number of
light flavours in the running coupling constant and in 
${\cal C}_{1,q}^{\rm NS}$ on the righthand side of Eq. (\ref{eqn:20})
will be enhanced by one unit.


\begin{thebibliography}{99}
\bibitem{Wit} E. Witten, Nucl. Phys. {\bf B104} (1976) 445;\\
J. Babcock and D. Sivers, Phys. Rev. {\bf D18} (1978) 2301;\\
M.A. Shifman, A.I. Vainstein and V.J. Zakharov, Nucl. Phys. {\bf B136}
(1978) 157;\\
M. Gl\"uck and E. Reya, Phys. Lett. {\bf B83} (1979) 98;\\
J.V. Leveille and T. Weiler, Nucl. Phys. {\bf B147} (1979) 147.
\bibitem{Wat}
D.A. Watson, Z. Phys. {\bf C12} (1982) 123.
\bibitem{Bmsmn}M. Buza, Y. Matiounine, J. Smith,
R. Migneron and W.L. van Neerven, Nucl. Phys. {\bf B472} (1996) 611.
\bibitem{Bmsn1}M. Buza, Y. Matiounine, J. Smith,
and W.L. van Neerven, Nucl. Phys. {\bf B485} (1996) 420.
\bibitem{Lrsn} E. Laenen, S. Riemersma, J. Smith and W.L. van Neerven,
Nucl. Phys. {\bf B392} (1993) 162, 229,\\
S. Riemersma, J. Smith and W.L. van Neerven, Phys. Lett.
{\bf B347} (1995) 143,\\
B.W. Harris and J. Smith, Nucl. Phys. {\bf B452} (1995) 109.
\bibitem{Bmsn2} M. Buza, Y. Matiounine, J. Smith,
and W.L. van Neerven, DESY 96-278, hep-ph/9612398, to be published in
Z. Phys. {\bf C} (1997).
\bibitem{Buza} M. Buza et al., "Deep Inelastic Production of Heavy Quarks",
Proceedings of the Workshop "Future Physics at HERA", DESY, Hamburg, 25-26
September 1995, p. 393-401.
\bibitem{Acot}
M.A.G. Aivazis, J.C. Collins, F.I. Olness and W.K. Tung,
Phys. Rev. {\bf D50} (1994) 3102.
\bibitem{Neer}
W.L. van Neerven, "Charm Production in deep Inelastic Lepton-Hadron Scattering"
, Lectures presented at the XXXVIIth Cracow School of Theoretical Physics, 30th
May- 10th June, Zakopane, Poland. To be pub. in Acta Phys. Polon. B.
\bibitem{Grsv}   M. Gl\"uck, E. Reya, M. Stratmann and W. Vogelsang,
Phys. Rev. {\bf D53} (1996) 4775.
\bibitem{H1}C. Adloff et al. (H1 Collaboration), Z. Phys. {\bf C72}
(1996) 593.
\bibitem{Zeus}M. Derrick et al. (ZEUS Collaboration), Phys. Lett. {\bf B349}
(1995) 225;\\
M. Derrick et al. (ZEUS Collaboration), XXVII Int. Conf. on HEP '96, Warsaw
(1996);\\
J. Breitweg et al. (ZEUS Collaboration), DESY-97-089.
\bibitem{Lave} S.A. Larin and J.A.M. Vermaseren, Phys. Lett. {\bf B259} (1991)
345.
\bibitem{Sam} M.A. Samuel, J. Ellis, M. Karliner, Phys. Rev. Lett. 74 (1995)
 4380.
\bibitem{Blum} J. Bl\"umlein and W.L. van Neerven, to appear.
\end{thebibliography}
\end{document}